# Stabilizing a novel hexagonal Ru$_2$C through Lifshitz transition under pressure

Weiwei Sun[a,b], Sudip Chakraborty[a,b], Klaus Koepernik[c], Rajeev Ahuja[a,b]

a Applied material physics, Department of material science and engineering, KTH-Royal Institute of Technology, 10044 Stockholm, Sweden

b Condensed Matter Theory Group, Department of Physics and Astronomy, Uppsala University, SE-751 20 Uppsala, Sweden;

c Leibniz Institute for Solid State and Materials Research, IFW Dresden, D-01069 Dresden, Germany

## Abstract

A new type of heavy transition metal carbide (TMC), Ru$_2$C with a space group of $p\bar{3}m1$(164) was synthesized experimentally at high pressure-high temperature [J Phys. Condens. Matter. 2012 Sep 12; 24(36): 362202.] and it was consequently quenched to ambient condition. We have carried out the dynamical stability study, which reveals the instability at ambient condition. The effect of pressure has been taken into consideration in order to stabilize as the reported synthesizing condition. We have found that it can be stabilized from 30 GPa to 110 GPa. The stronger $4d$-$2p$ hybridization and the formation of a cage like Fermi surface do impact the stability and also illustrates a Lifshitz transition. We have also found a mixed $4d$-$2p$ bands crossing the Fermi level form a Fermi surface piece at $\Gamma$ point under pressure. The freshly appearing bands provide a tunnel for quantum transportation and it reduces the density of states at Fermi level, which further stabilizes the lattice under pressure.

## Introduction

A series of materials with superior mechanical properties like hardness has been one of the long-standing research explorations in material science community. The super hard material cubic boron nitride (c-BN) can be synthesized under high-pressure conditions. The light covalent elements like B, C, N, O and valence electron-rich transition metals like Ta, W, Re, Os, Ru are known to be combined in the synthesis process [1-3] to explore new type of hard materials. Particularly, transition metal (TM) carbides and nitrides are considered as potential candidates for industrial applications due to their hardness, high melting point, outstanding thermal conductivity and reasonable corrosion resistance [4].

The above approach essentially depends on the well-known formation of *p-d* hybridized covalent bond between light covalent elements and TMs. It is observed that formation of carbides and nitrides of latter TMs are not so favorable due to their formation energies compared with lighter TMs [5,6,7]. The applied external pressure is capable of changing the physical and chemical properties of materials and enhancing their chemical activities. The successful high-pressure syntheses of $Re_2C$ [8], $OsN_2$ [9] and $Re_2N$ [10] have been reported recently and $Ru_2C$ has been synthesized successfully under both high pressure and temperature [11], where it is quenched to room temperature at its equilibrium. The simulated XRD pattern has a good agreement with the experimental data in the corresponding finding, but the rare appearance of space group of $p\bar{3}m1(164)$ of latter TM carbides at low temperature and the difference between computed and experimental lattice parameters becomes the motivating fact of this work. The profound comparative analysis of Ru, Os, Re and other transition metal carbides and nitrides will be quite intuitive to give an insight on the possible crystalline structure. $Re_2N$ and $Re_2C$ having space group P63/mmc (194) [8,10] can only be stable at high temperature-high pressure condition. Both $Cr_2N$ and beta-$Ta_2N$ [12,13] show space group of P63/mmc (194) at ambient condition and $Fe_2N$ and $Mo_2C$ shows stability in the space group of Pbcn (60) at ambient condition while they are showing $p\bar{3}m1(164)$ symmetry only at high temperature [14]. None of them crystallizes in a space group of $P\bar{3}m1$ (164) at ambient condition which opens up to explore the hydrostatic quenching process more extensively. Investigating the dynamical stabilty at ambient condition can clarify this phenomena to a great extent. The imaginary modes in phonon dispersion curve (PDC) from ambient condition to 110 GPa can be conclusive regarding its stability scenario. Due to the experimental condition of high pressure in elevated temperature, studies of electronic temperature effect on the stability have been considered and it is reported [15,16] that the electronic temperature enables to stabilize the lattice but the applied temperature value is supposed to be much higher than the experimental temperature. Therefore, the electronic temperature effect has been discarded in this work and the influence of pressure is investigated. The electronic structure calculations reveal the strength of the 4d-2p hybridization and the density of states at Fermi level with applied external pressure. A cage like Fermi surface (FS) piece has been evolved at Γ point consequently. In order to investigate it more deeply, we have performed the band structure analysis, which indicates the possibility of Ru-C anti-bonding states reversing the bonding states that cross Fermi level at Γ point. This work will provide a profound insight regarding the possible stable configuration range for heavy transition metal carbides. Their unique characteristics pave the way for an extensive analysis of its phase diagram, which is quite significant from the industrial applications aspect. This work can be a motivating one from the perspective of experiments at ambient condition.

# Computational methods

To calculate the relevant quantities of $Ru_2C$, the density functional theory (DFT) calculations were carried out with Vienna Ab-initio Simulation Package (VASP)[17,18] and full-potential local-orbital code (FPLO9.01.35)[19]. In VASP, generalized gradient approximation (GGA) [20] in PBE functional form was used which incorporated the formalism of projected-augmented wave (PAW) method [21] in order to relax the structure and determine the force constant. For the convergence of physical quantities, a mesh of 13 × 13 × 9 k-points was used. For Brillouin Zone integration, the first order of Methfessel-Paxton method [22] with a 0.2 eV width was applied for the smearing, which obtains the maximum entropy value of 1 meV per atom. Relaxation convergence of 0.01 eV for ionic positions, 0.001 meV for electronic convergence, cut-off energy of 650 eV were employed throughout the calculations. Within PAW, $5d^4$ and $6s^1$ states were treated as valence states for Ta; $2s^2$ and $2p^2$ for C. The structures were optimized to equilibrium by relaxing both the cell shape and the internal atomic coordinates when its symmetry was preserved. For testing the dynamical stability of the system, Phonopy [23] was employed within Density Functional Perturbation Theory (DPFT) implemented in VASP. In DPFT calculations, a 4 × 4 × 3 super cell consisting of 144 atoms meshed by 5×5×5 k-points was constructed to calculate phonon dispersion relation. To achieve a reasonable convergence, an energy cutoff of 700 eV and $10^{-8}$ energetic precision were pursued. FPLO was used to calculate the electronic structure based on the VASP relaxed lattice. Within FPLO, PBE as an exchange correlation functional in GGA within four-component full relativistic framework was employed to investigate the electronic structure extensively. All the calculations were carefully converged with respect to the regular k-point mesh and 25×25×25 were found to be the most feasible. The expanded valence basis was consisted of *4s, 4p, 5s, 6s, 4d, 5d, 5p* states in the case of Ru; *1s, 2s, 2p, 3s, 3p, 3d* states for C.

# Results and Discussion

## A. Dynamical stability under pressure

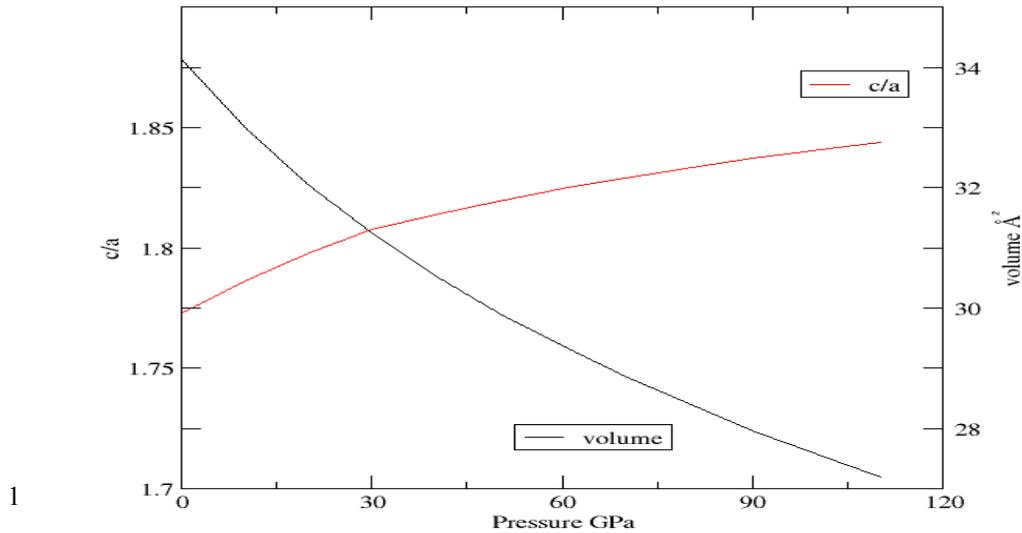

Fig. 1 the relation of the unit cell volume and c/a ratio variation with pressure

We have shown *P - V* & *P - c/a* relations in Fig.2. To observe the behavior of lattice changing under pressure, 10 GPa is chosen as interval up to 110 GPa. As the *P - V* curve shows, the volumes decrease smoothly and the c/a values are continuously increasing with pressure, which reveals that the lattice is elongated along Z direction by compression. The computed lattice constants and equilibrium volume at ambient condition are a = 2.81Å, c = 4.98 Å and $V_{eq}$= 34.14 Å$^3$. Our calculated results are very closed to the results in Ref. [11]. The simulations of the off stoichiometric effects on lattice constants show 10–15% difference in the Re–C system [24] and the bulk moduli is underestimated in Os-c systems [25,26]. However, the actual physical reason is still unexplored. We should mention that the internal coordinates are relaxed as well to achieve a force convergence. From the symmetry point of view, Ru atoms (2d wyckoff position) are variable instead of fixed 1/4. The discrepancy in the lattice parameters does motivate us to investigate the stability at equilibrium.

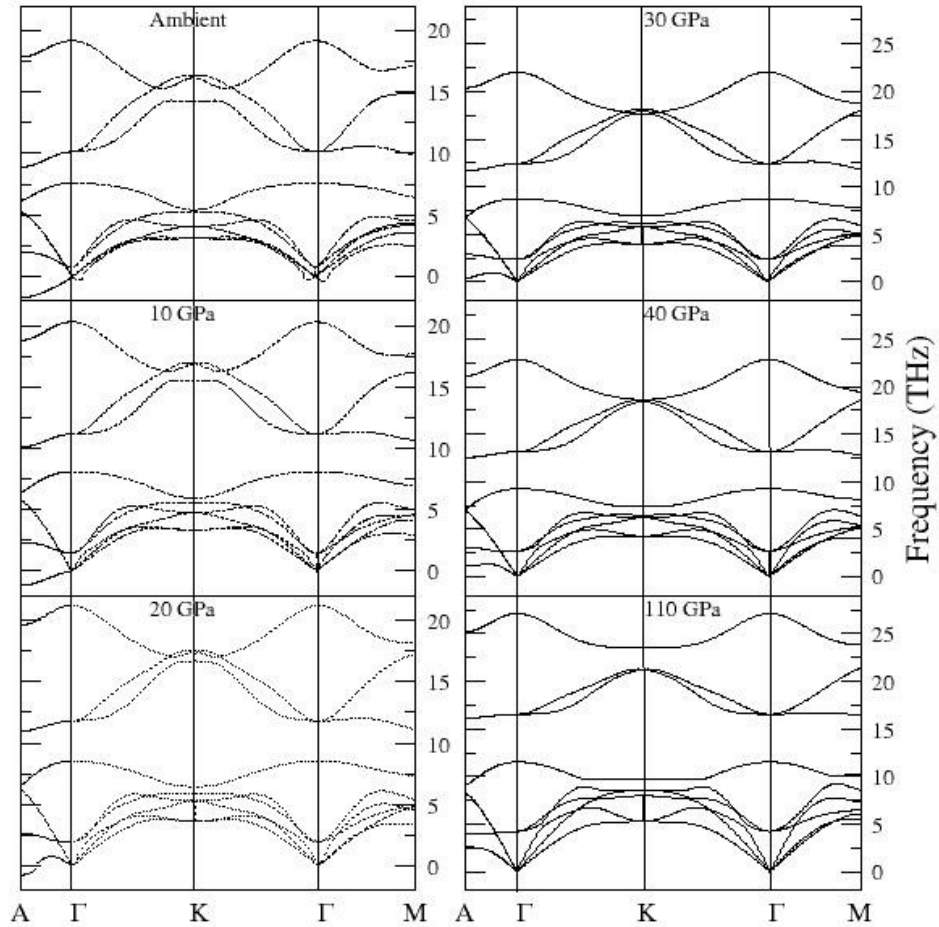

Fig.2 the phonon dispersion curves (PDCs) at 0GPa, 10 GPa, 20 GPa, 30 GPa, 40 GPa and 110 GPa

In Fig.2, PDC has been shown along A – Γ – K – Γ – M direction from 0 GPa to 110 GPa. The imaginary modes have been found along A- Γ direction, namely (00ξ) direction and some less negative modes along K – Γ – M that are in the (ξξ0) direction at ambient condition and hence $Ru_2C$ is unstable at ambient condition. But whether the pressure can stabilize the lattice is required to be answered. As the compression increases, the imaginary modes turn to be positive from 30 GPa. The positive phonon modes are climbing to higher frequency with pressure up to 110 GPa. $Ru_2C$ with the space group of $P\bar{3}m1$ (164) is unstable at ambient condition but it is stable in the range from 30 GPa till 110 GPa. The wide stable pressure region paves the way for potential application in extreme condition.

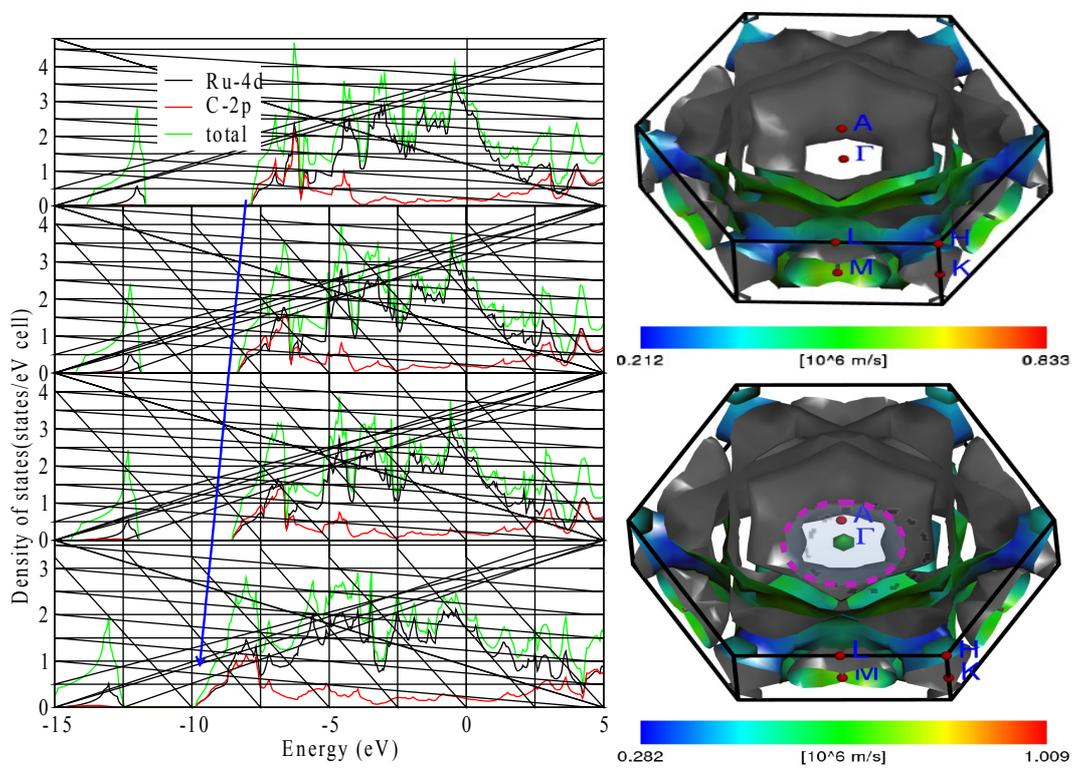

## B. Electronic structure

Fig. 3 The lefe panel represents the total and projected DOS of Ru2C at ambient, 20 GPa, 30 GPa and 110 Gpa from top to bottom. The Fermi surfaces at ambient (top) and 30 GPa(bottom) are located at the right panel.The dotted blue line in DOS represents the broadening of hybridizied peak. The purple dotted ring circulates the emgergence of the FS piece at the zone center.

The total density of states (DOS) and projected DOS of $Ru_2C$ under different pressures are shown in Fig.3. In our previous discussion, $Ru_2C$ at ambient condition is unstable which is verified by PDC while it is stabilized from 30 GPa. We know that the *d-p* hybridization is the typical characteristic in TMCs and DOS under pressure can govern the stronger chemical bond. Besides, the pressure has shifted the DOS to the deeper energy region (marked by the dotted blue line) and the hydride peaks become broad, which shows a sign of stronger Ru – C bond. The decreasing of the bond length and the stronger hybridization harden the lattice under pressure. Particularly, the states at Fermi level address our focus due to the metallic nature of the system. Ru 4d states are the dominant states and they are quite high compared with the cases under pressure. Additionally, the states at Fermi level fall down from 82.21 states/Ha to 64.84 states/Ha per cell from ambient pressure to 30 GPa. This reduction of number of states driven by pressure stabilizes the lattice. In a nutshell, this interesting discovery of the shift at the Fermi level pushes us to investigate the Fermi surface (FS) behavior more explicitly in future. The obtained shape of FS shows its complexity and the evident difference lies at the Γ point, where a cage like piece appears under pressure. We should mention here that no other noticeable changes have been found by any single band crossing Fermi level at Γ point. This cage is expanding with pressure showing a high momentum. This phenomenon is reasonably connected with the band structure description, particularly existence of the partially occupied bands formed by compression. The following analysis is particular on the band structure with the weighted C-*2p* orbital (red) in Fig. 4.

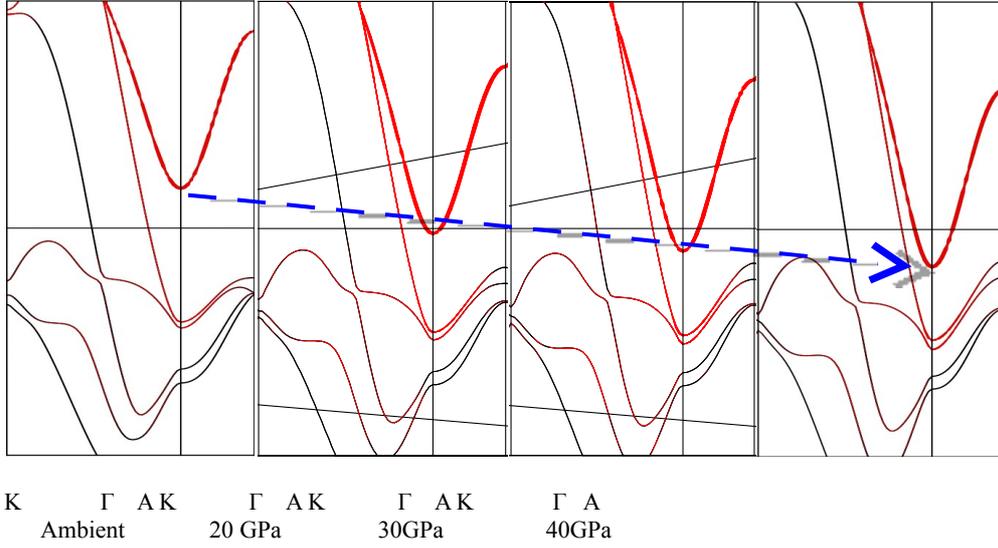

Fig.4 The band structure along K-Γ-A (the direction is from xy plane to z direction) at ambient, 20 GPa, 30 GPa and 40 GPa with the weighted C-*2p* orbital(red). The dotted line is heading towards the bottom of the weighted band.

Specifically one can see the signature of mixed Ru-4d band and the intense band penetrates into the valence band region at Γ point. The newly shaped band forms the cage like FS piece as shown in Fig. 3 providing a charge transfer tunnel between conduction band and valence band. The electronic motion can be carried out smoothly through this tunnel, which contributes to the less number of states at Fermi level. At the beginning, the net valence population analysis (neglecting hybridization with Ru) was performed to search for the potential targets. For Ru, rather tiny shift of occupation on a specific orbital has been uncovered by comparing ambient and 30 GPa cases. The phenomenon is not been reproduced on C atom that gains 0.2 more valence electrons (primarily from *2p* orbital) compared with that in equilibrium. For the gross valence number, the *2p* orbital valence number only differ 0.04, which shows the reservation of the total valence electrons and the more *2p* electrons get very closed to the core region, which aren't like to participate chemical bonding as shown in Fig.3 where the *4p* – *2p* hybridized approaches to the deeper energy level. Moreover, the FS piece in the zone center shows a high Fermi velocity indicating efficient quantum transportation through this band. The transportation tunnel crossing Fermi level, which has been created by the FS piece, provides the higher probability of the motion of electrons reduces the states at the Fermi level, which indicates the stability of Ru$_2$C under pressure. In 1960, Lifshitz had predicted that anomaly occurs if the topology of the Fermi surface changes under pressure while keeping the number of valence electrons constantly [27]. In the case of Ru$_2$C, the topology of the Fermi surface clearly changes when $4d - 2p$ DOS peak gets delocalized (see Fig. 3) as an example of Lifshitz transition. Two more bands at Γ point cross Fermi level reflects the change of symmetry and corresponds the emergence of the new piece as shown in Fig. 4. The low occupancy of the bands at Fermi level uncovers that the established tunneling manifests a minor influence on the stabilizing under pressure. The reduction of states at Fermi level is result from the downshift of DOS and the quantum tunneling formed under pressure, which also drives the mixed $4d - 2p$ bands deeply. The downshift and broadening of $4d - 2p$ further strengthen the chemical bonds and the exotic change of topology of FS also verifies Lifshits transition. Essentially, the finding of Lifshitz transition fruits the physics of TMC, which open a door to the versatile phase transitions.

# Conclusion

The structure of Ru$_2$C with a space group of $p\bar{3}m1(164)$ is unstable at ambient condition but it can be stable from 30 GPa till 110 GPa at 0K. In its equilibrium, the softening phonon modes mainly lie in Z direction and they are enhanced in positive direction along with the external pressure from 30 GPa. The pressure weakens the $4d - 2p$ hybridization and broadens it in the range of -7 eV to -10 eV. Contrary to the ambient condition, two more bands cross Fermi level and they form a cage like FS piece at Γ point, which shows a Lifshitz transition. This tunnel connecting the conduction band and valence band shows a high Fermi velocity, which can explain the highly efficient quantum transportation. In addition to this, the bottom of the band falling down monotonically, which provides a transportation tunnel to reduce the occupied states at Fermi level consequently. The mixed bands downshifting towards to the core and the broadening $4d - 2p$ hybridization reveals the underlying the stability under external pressure. The exotic phenomenon on the density of states and Fermi surface is attribute to the stronger hybridization of Ru $4d$ – C $2p$, which stabilizes the lattice.

Acknowledgement

We are grateful to HPC2N in Sweden for the computational power and the fruitful discussions with Dr. Manuel Richter in IFW Dresden. Weiwei Sun acknowledges the CSC scholarship for financial support and ESF Progamme "Advanced Concepts" (Psi-k2) for travel grant.